\documentclass{article}
\usepackage{arxiv}
\usepackage{amsmath}
\usepackage[utf8]{inputenc} 
\usepackage[T1]{fontenc}    
\usepackage{hyperref}       
\usepackage{url}            
\usepackage{booktabs}       
\usepackage{amsfonts}       
\usepackage{nicefrac}       
\usepackage{microtype}      
\usepackage{lipsum}		
\usepackage{graphicx}
\usepackage{natbib}
\usepackage{doi}
\usepackage{chemfig}
\usepackage{subcaption}
\usepackage[newcommands]{ragged2e}
\usepackage{graphicx, caption}
\usepackage{dcolumn}
\usepackage{bm}

\title{Advancements and Applications of NMR and MRI Technologies in Medical Science: \\A Comprehensive Review}

\author{ \href{https://orcid.org/0000-0002-4924-3033}{\includegraphics[scale=0.06]{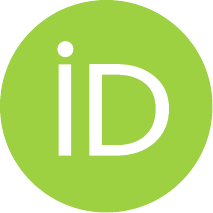}\hspace{1mm}Islam G. ~Ali}\thanks{islam.gamal@sci.aru.edu.eg} \\
Physics Department\\
Faculty of Science\\ Arish University, Egypt\\}

\renewcommand{\shorttitle}{\textit{arXiv} Template}

\hypersetup{
pdftitle={A template for the arxiv style},
pdfsubject={q-bio.NC, q-bio.QM},
pdfauthor={David S.~Hippocampus, Elias D.~Striatum},
pdfkeywords={First keyword, Second keyword, More},
}

\begin{document}
\maketitle

\begin{abstract}
	Nuclear Magnetic Resonance (NMR) and Magnetic Resonance Imaging (MRI) represent versatile tools with diverse applications spanning physics, chemistry, geology, and medical science. This comprehensive review explores the foundational principles of NMR and MRI technologies, elucidating their evolution from fundamental quantum mechanical concepts to widespread applications in medical science. Commencing within a quantum mechanical framework, the concise review emphasizes the significant role played by NMR and MRI in clinical research. Furthermore, it provides a succinct survey of various NMR system types. Conclusively, the review delves into key applications of MRI techniques, presenting valuable methodologies for visualizing internal anatomical structures and soft tissues.
\end{abstract}


\section{\label{sec:level1}Introduction}

Nuclear Magnetic Resonance (NMR) and Magnetic Resonance Imaging (MRI) technologies represent pivotal tools in the scientific arsenal, facilitating a comprehensive understanding of molecular structures and providing intricate visualizations of internal anatomical features within the human body. NMR, distinguished as a non-destructive analytical technique, harnesses nonionizing radio frequency (RF) waves to meticulously characterize the chemical composition of both living and non-living entities. Conversely, MRI stands as a cornerstone in medical science, widely embraced for diagnostic imaging applications \cite{sakurai1995modern, griffiths2018introduction, feynman2010quantum, khashami2024fundamentals}.

The genesis of NMR and MRI technologies is inextricably linked to the unfolding revelations within the atomic world and the tenets of quantum theory. A profound comprehension of these technologies necessitates a delve into the intricacies of quantum physics, particularly the foundational concept of spin \cite{maleki2020spin, maleki2021quantummetrology, maleki2021quantumphase}.

The exploration of quantum spin's discovery stands out as one of the most captivating narratives in the annals of modern science. Its origins trace back to a deeper comprehension of matter, brought into focus by unraveling the constituents of atoms. This journey commenced with Rutherford's revelation of the atomic nucleus in 1911, followed by Bohr's model of the atom in 1913 \cite{hughes1990bohr, rutherford1932discussion}. Subsequently, de Broglie's proposition of wave-particle duality confronted classical physics paradigms, triggering a profound transformation in the landscape of quantum mechanics. This departure from classical viewpoints imbued quantum systems with counterintuitive behaviors, fueling ongoing debates \cite{maleki2023revisiting, maleki2021quantumeraser, maleki2019stereographic}.

The paradigm shift introduced by quantum mechanics offered a novel perspective on the physical realm, characterized by probabilistic outcomes, superposition, and particle entanglement \cite{khashami2013entanglement, maleki2021natural}. This redefined worldview continues to evoke discussions and scrutiny, contributing to the enduring intrigue surrounding the unconventional behavior of quantum systems.

The formalization of quantum mechanics transpired through Heisenberg's mechanics in 1925 and Schrödinger's wave equation in 1926. Concurrently, Uhlenbeck and Goudsmit uncovered electron spin and its intrinsic angular momentum properties, propelling Paul Dirac to revise the hydrogen atom's spectrum in 1928. The inception of NMR technology, marked by Rabi's discovery in 1938, was further propelled in 1946 when Felix Bloch and Edward Purcell developed the first NMR machines. These milestones were intricately entwined with the understanding of the Zeeman effect and the interaction of nuclear spins with external magnetic fields in the development of quantum mechanics \cite{bloch1946nuclear, friston1994analysis, roberts1977basic}.

The elucidation of the Bloch equation played a pivotal role in deciphering the dynamics of the NMR system, offering insights into spin relaxation dynamics within a quantum spin system. The journey continued with the description of the NMR signal using the Fourier transform (FT) method, converting the time-domain NMR signal into the frequency domain signal. Richard R. Ernst's advancements culminated in the development of FT-NMR spectroscopy in 1991 \cite{feynman2010quantum, butz2006fourier, rabi1938new}.

It is noteworthy that NMR and MRI, representing distinct generations of Rabi's initial discovery, have evolved from theoretical concepts into practical tools with transformative applications. These technologies have ushered in a new era of medical imaging, offering diagnostic capabilities for investigating both pathological and normal tissues \cite{graham1985anaerobic, abragam1983principles, solomon1955relaxation}.

This comprehensive review directs its focus towards unraveling the quantum mechanical foundations underpinning NMR and MRI. By attempting to construct an accessible foundation for the quantum mechanical framework of these technologies, the review aims to bridge the gap between technological applications and the fundamental principles that govern their functionality. Understanding these basic frameworks holds immense value for numerous reasons \cite{khashami2024fundamentals, khashami2021tracking}.

\section{Zeeman Effect}

\begin{figure}[h]
     \centering
     \begin{subfigure}[b]{0.3\linewidth}
         \centering         \includegraphics[width=0.7\linewidth]{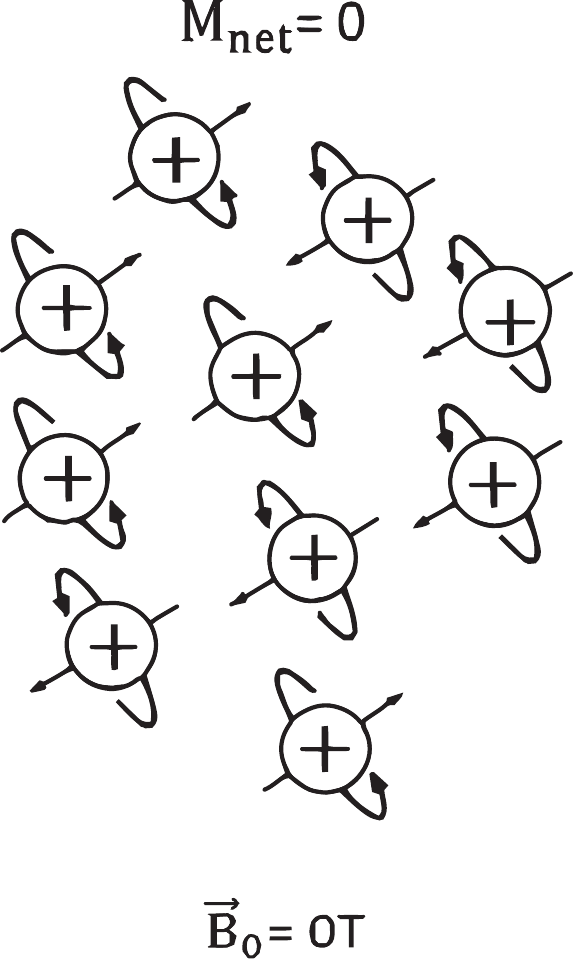}
         \caption{}
         \label{fig.1a}
     \end{subfigure}
     \quad
      \begin{subfigure}[b]{0.4\linewidth}
         \centering        \includegraphics[width=\linewidth]{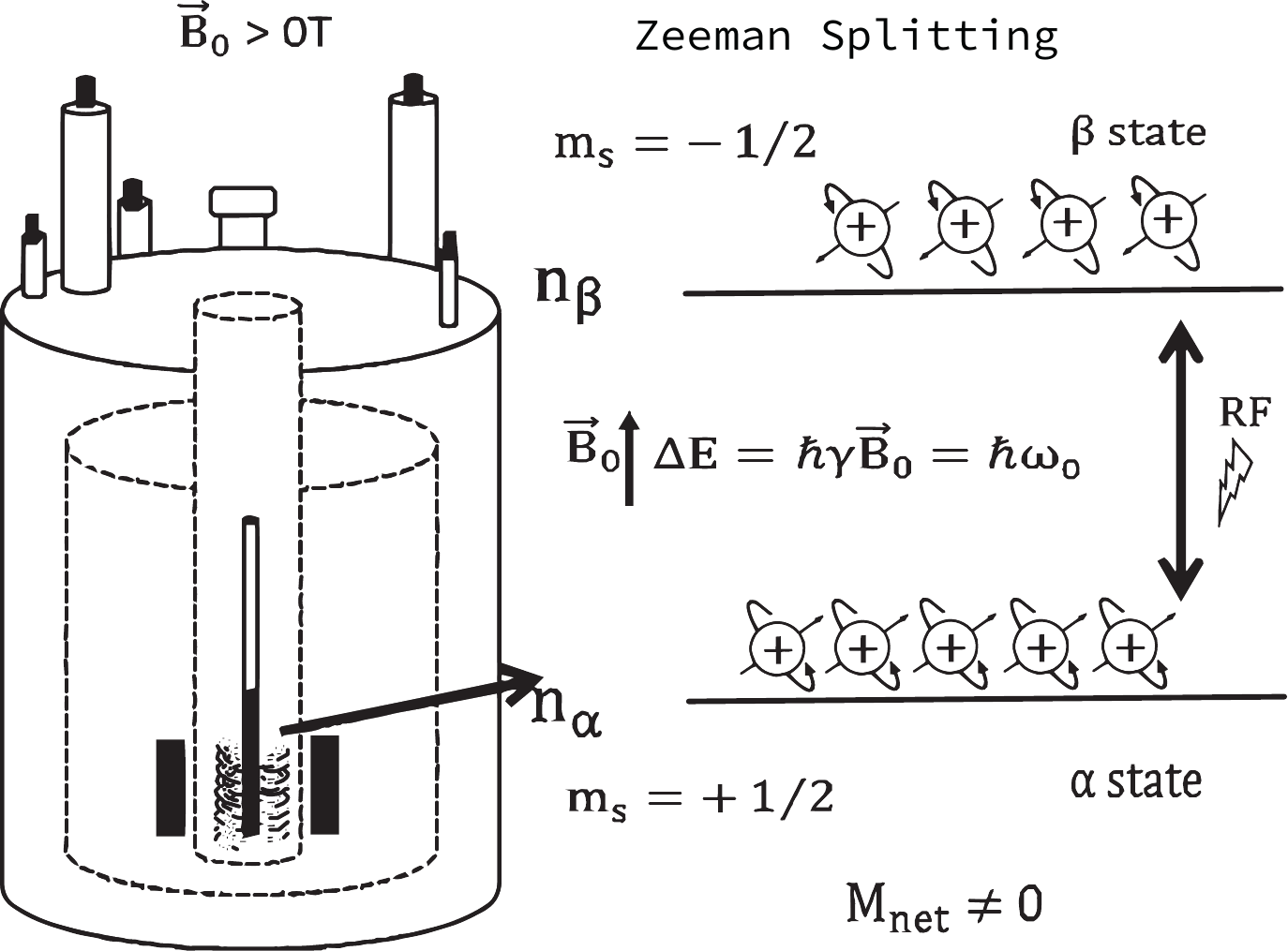}
         \caption{}
         \label{fig.1b}
     \end{subfigure}
     \quad
     \begin{subfigure}[b]{0.3\linewidth}
         \centering        \includegraphics[width=\linewidth]{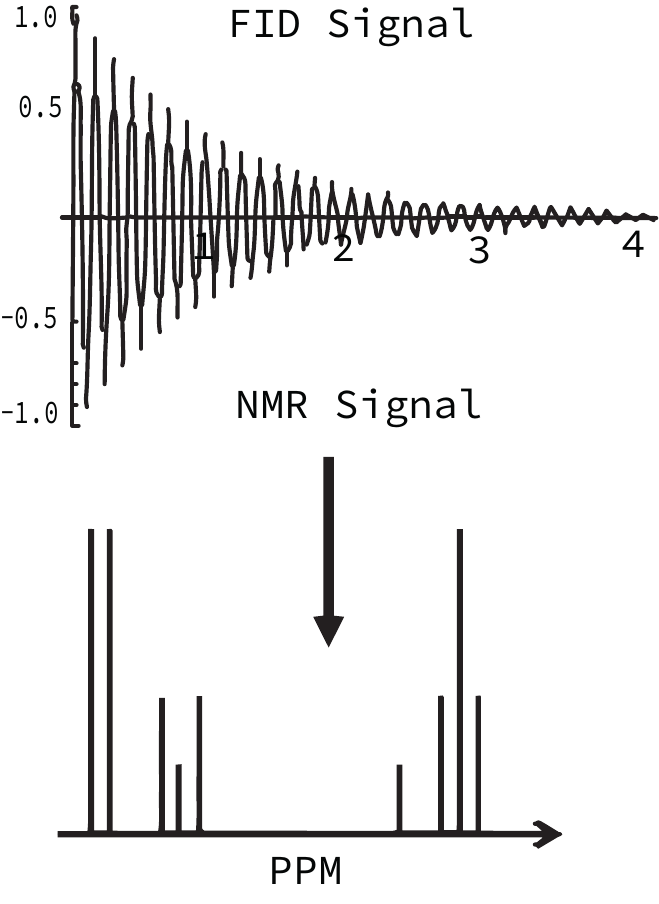}
         \caption{}
         \label{fig.1c}
     \end{subfigure}
      \quad
      \captionsetup{justification=Justified}
    \caption{\label{fig.1}(a) In the absence of the magnetic field $\vec{B}_0$, $^{1}\text{H}$ atoms exhibit random orientations, and their nuclei are distributed across a single energy level. (b) Zeeman splitting occurs when a strong magnetic field is applied, causing some nuclei to align parallel to $\vec{B}_0$ ($\alpha$ state) and others anti-parallel to $\vec{B}_0$ ($\beta$ state). A larger proportion of the nuclei occupy the lower energy level. Following the application of an RF pulse, certain nuclei in the lower state transition to the higher state. (c) Upon deactivating the RF pulse, the nuclei return to equilibrium, emitting a free induction decay (FID) signal. Through Fourier transformation (FT), the NMR spectrum is derived from the FID signal.}
\end{figure}

To unveil the fundamental principles behind NMR and MRI technologies, we explore the dynamics of a spin system in the presence and absence of a magnetic field \(\boldsymbol{\vec{B_0}}\). Initially, spins possessing magnetic dipole moments \(\boldsymbol{\mu_s}\) align randomly, resulting in a net magnetic moment of zero, as depicted in Fig. \ref{fig.1a} \cite{abragam1983principles, khashami2021tracking, mansfield1977multi}. Upon the application of an external magnetic field \(\boldsymbol{\vec{B_0}}\), nuclei precess coherently, aligning either with or against the magnetic field based on their spin, as illustrated in Fig. \ref{fig.1b} \cite{maleki2023revisiting, emsley2013high}. The magnetic field induces a torque on \(\boldsymbol{\mu_s}\), aligning it along the z-axis. This alignment results in Zeeman splitting and gives rise to (2$I$ + 1) energy levels expressed by the spin quantum number \(m_s\) \cite{khashami2024fundamentals, davies1989nuclear, ames2010theory, lancaster2014quantum, maleki2016generation}.

From the Zeeman effect, the NMR signal relies on the energy difference between the two-level states of the nucleus. The lower energy level, the \(\alpha\) state (\(m_s = +1/2\)), aligns with the magnetic field, while the higher energy \(\beta\) state (\(m_s = -1/2\)) aligns against \(\mathbf{B_0}\). Typically, the \(\alpha\) state has slightly more occupants than the \(\beta\) state \cite{bloch1946nuclear, rabi1938new, ames2010theory, purcell1946resonance}.

The spin angular momentum in the z-axis is \(S_z = m_s \hbar = \pm \hbar/2\), and the total magnitude spin angular momentum vector for spin quantum number \(1/2\) is \(|\mathbf{S}| = \sqrt{s(s + 1)}\hbar = \sqrt{3}/4 \hbar\), where \(\hbar\) is the Planck constant (\(1.054 \times 10^{-34}\) J$\cdot$s). The magnetic dipole moment of nuclei, \(\mu_z\) with \(m_s = \pm 1/2\), is defined as \(\mu_z = m_s \hbar \gamma = \pm 1/2 \hbar \gamma\), where \(\gamma\) is the gyromagnetic ratio with units MHz/Tesla, a constant value for each nucleus, and is related to the magnetic moment \(\mu_s\) and the spin number \(I\) for a specific nucleus, expressed as
\begin{equation}
    \gamma = \frac{2\pi \mu_z}{\hbar I} \label{eq:gyromagnetic_ratio}
\end{equation}
This gyromagnetic ratio can be positive or negative \cite{khashami2024fundamentals, damadian1980field, kitaev2002classical, de1970reinterpretation}.

Due to Zeeman splitting, the interaction energy of a spin state with a magnetic field along the z-axis is proportional to the magnetic field strength and is expressed as
\begin{equation}
    E = {\vec \mu} \cdot {\vec{B_0}}
    \label{eq:interaction_energy}
\end{equation}
The energy along the z-axis is given by \(E = -\mu_z B_0\). Therefore, the energy difference between the two upper and lower energy levels is obtained as
\begin{equation}
    \Delta E = E_{\beta} - E_{\alpha} = \hbar \gamma \vec{B_0} = \hbar \omega_0 \label{eq:energy_difference}
\end{equation}
In this relation, the nucleus spin precesses at a specific frequency \(\omega_0\), known as the Larmor frequency, with the unit rad/s. As shown in Eq. (\ref{eq:energy_difference}), the Larmor frequency, the precession frequency of the spins around the axis of the magnetic field, is \(\omega_0 = \gamma \vec{B_0}\) \cite{khashami2024fundamentals, jackson1999classical}. Resonant RF pulses, typically in the frequency range of 60-900 MHz, can be utilized to excite nuclei in the lower energy state \cite{abragam1983principles, mansfield1977multi, damadian1980field, lauterbur1973image, bushong2013magnetic, hashemi2010basic}. Upon turning off the RF pulse, nuclei precess around \(\boldsymbol{\vec{B_0}}\) \cite{hashemi2010basic, heisenberg1989encounters}.

Zeeman splitting results in a net magnetization of the sample, causing nuclei to transition to a lower energy state when the magnetic field is turned off. The rotating magnetization vector induces a current in a receiver coil, facilitating the detection of the NMR signal. The oscillating signal manifests as waves with decreasing magnitude as the nuclei realign with the magnetic field. This phenomenon, known as the free induction decay (FID) signal, is then transformed to the frequency domain using the Fourier transformation (FT) method, presenting as peaks along the x-axis of the NMR spectrum. This phenomenon, referred to as chemical shift, is expressed in units of parts per million (ppm) [ refer to Fig. \ref{fig.1c}] \cite{solomon1955relaxation, hashemi2010basic}. The properties of chemical shift will be further elucidated later in this review.

\section{The Boltzmann Distribution}

The Boltzmann distribution characterizes the distribution of nuclei in different spin states. At room temperature, the number of spins in the lower energy level (\(n_\alpha\)) surpasses those in the higher energy level (\(n_\beta\)). However, the energy gap between these states is relatively small, making the distribution temperature-dependent \cite{khashami2024fundamentals, solomon1955relaxation, bushong2013magnetic, maleki2015entanglement}.

The Boltzmann distribution dictates that NMR signal intensity is directly proportional to the applied magnetic field and inversely proportional to temperature. The weak magnetic moments of nuclear spins lead to minor variations in nuclear spin populations \cite{levitt1982broadband, schwabl2013statistical, maleki2019quantum}. In essence, the NMR signal correlates with the nuclear polarization (\(P\)), determined by the Boltzmann distribution of nuclear spins across Zeeman energy levels. In thermal equilibrium, for a spin-1/2 system, it is influenced by the thermal population difference described by the Boltzmann factor as
\begin{equation}
  P = \tanh\left(\frac{\gamma\hbar B_0}{2k_BT}\right) \label{eq:spin_polarization}  
\end{equation}

where \(P\) is the spin polarization, \(T\) is the temperature, and \(k_B\) is the Boltzmann constant. The relative population of upper and lower states is given by \(n_\alpha + n_\beta = n_0\), where \(n_0\) is the number of nuclei per unit volume \cite{davies1989nuclear, stoneham1969shapes, levitt2013spin}.

At thermodynamic equilibrium, the ratio between the populations of the two levels can be expressed as
\begin{equation}
    \frac{n_\alpha}{n_\beta} = \exp\left(-\frac{\Delta E}{k_BT}\right) = \exp\left(-\frac{\hbar\omega_0}{k_BT}\right) \approx 1 - \frac{\hbar\omega_0}{k_BT} \label{eq:population_ratio}
\end{equation}

\noindent From Eq. (\ref{eq:population_ratio}), 
\begin{multline*}
    n_\beta = \frac{n_0 }{1 + e^{\beta\hbar\omega_0}} \approx \frac{n_0}{2 +{\beta\hbar\omega_0}} =  \frac{n_0}{2} \left( \frac{1}{1+\beta\hbar\omega_0/2}\right) \\ \approx \frac{n_0}{2}\left(1 - \frac{\beta\hbar\omega_0}{2}\right),
\end{multline*}
\begin{multline}
    n_\alpha = \frac{n_0 }{1 + e^{-\beta\hbar\omega_0}} \approx \frac{n_0}{2 -{\beta\hbar\omega_0}} =  \frac{n_0}{2} \left( \frac{1}{1-\beta\hbar\omega_0/2}\right) \\ \approx \frac{n_0}{2}\left(1 + \frac{\beta\hbar\omega_0}{2}\right) \label{eq:population_ratio_alpha}
\end{multline}

Furthermore, the magnetization for a spin-1/2 system can be calculated in terms of temperature (\(T\)) and the magnetic field (\(\boldsymbol{\vec{B_0}}\)). The total magnetization is expressed as \(M_0 = n_\alpha \mu_z^{(\alpha)} + n_\beta \mu_z^{(\beta)}\), representing the sum of all spins in the sample known as bulk magnetization. At equilibrium, the magnetization aligned with \(\boldsymbol{\vec{B_0}}\) is given by Curie's law for spin-1/2 nuclei as
\begin{equation}
    M_0 = (n_\alpha - n_\beta) \frac{\gamma\hbar}{2} \approx \frac{\gamma\hbar}{2} n_0 \frac{\hbar\omega_0}{2k_BT} = n_0 \frac{\gamma^2\hbar^2}{4k_BT} B_0 = \chi_0 B_0, \label{eq:magnetization}
\end{equation}
where \(\chi_0\) is the static nuclear susceptibility of a sample, indicating how the system magnetizes in \(\boldsymbol{\vec{B_0}}\) and is inversely proportional to \(T\).

\section{Chemical Shift}

\begin{figure}[h]
     \centering
     \begin{subfigure}[b]{\linewidth}
         \centering       \includegraphics[width=0.5\textwidth]{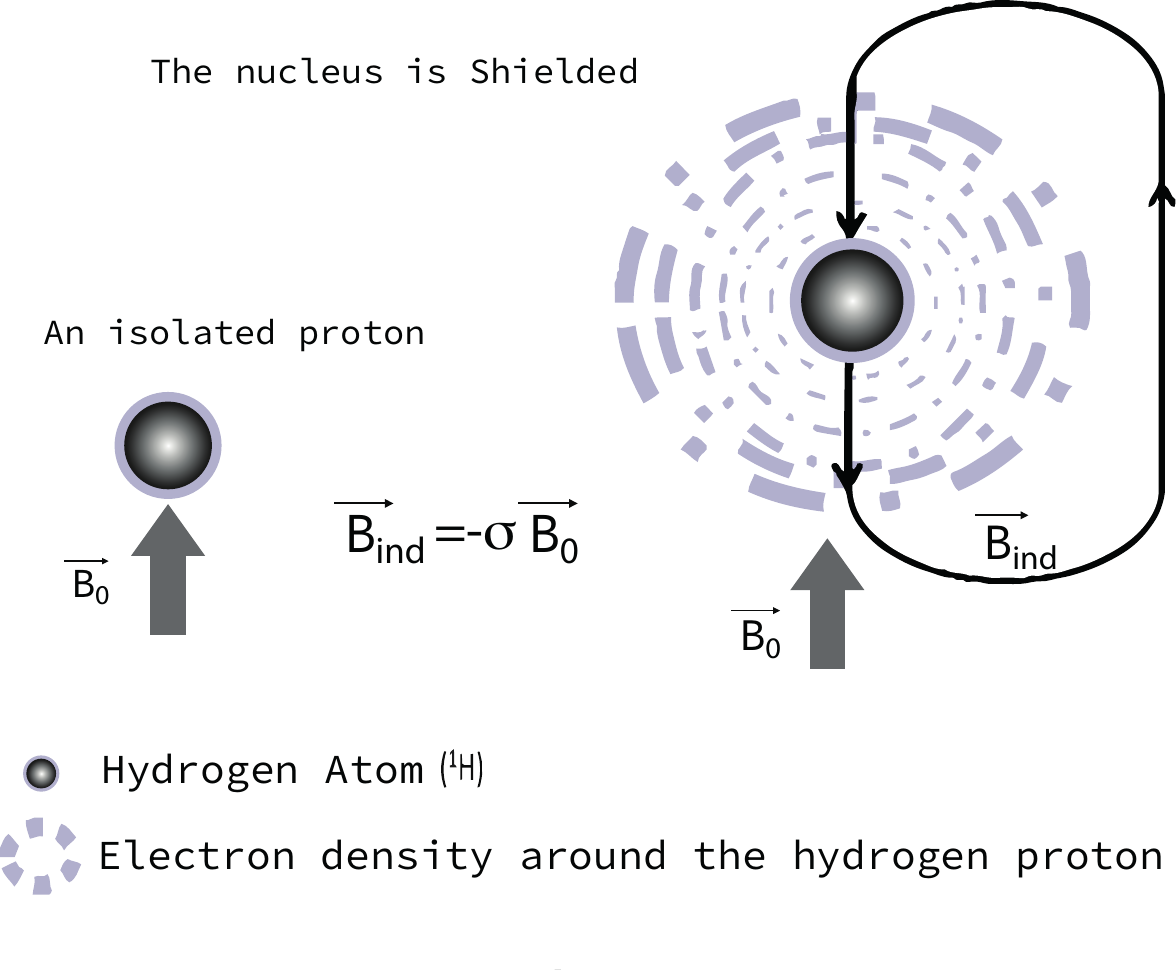}
         \caption{}
         \label{fig.2a}
     \end{subfigure}
     \quad
      \begin{subfigure}[b]{\linewidth}
         \centering
\includegraphics[width=0.5\textwidth]{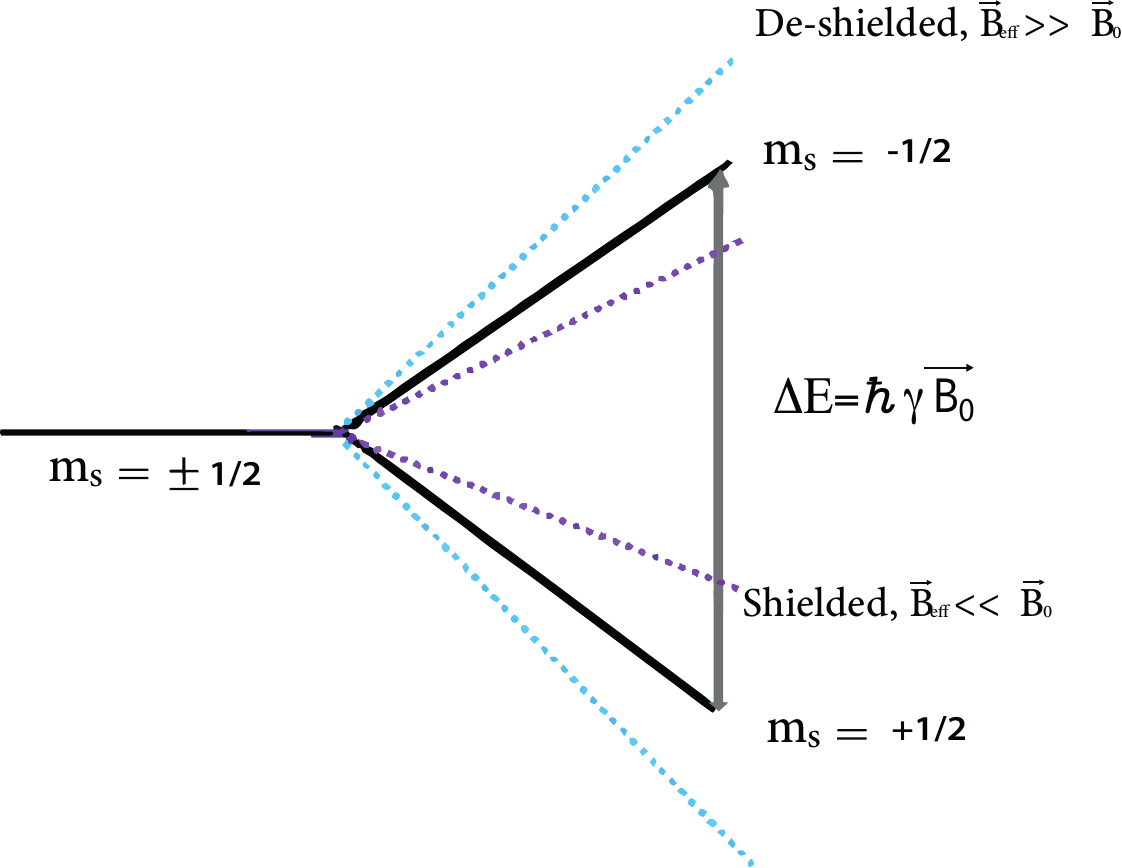}
         \caption{}
         \label{fig.2b}
     \end{subfigure}
      \quad \captionsetup{justification=Justified}
    \caption{\label{fig.2}(a) Within hydrogen nuclei, electrons induce alterations in the effective magnetic field ($\vec{B}_{\text{eff}}$) experienced by atoms. The external field ($\vec{B}_0$) generates currents around the nucleus, creating an induced magnetic field ($\vec{B}_{\text{ind}}$) opposing $\vec{B}_0$ and resulting in electron shielding of the nucleus. (b) With a spin quantum number of $1/2$, there exists a degeneracy of $(2I + 1)$ states, and the energy difference between these states is $\Delta E = -\hbar\gamma \vec{B}_0$. When $\vec{B}_{\text{eff}}$ is less than $\vec{B}_0$, the nucleus is shielded, whereas when $\vec{B}_{\text{eff}}$ exceeds $\vec{B}_0$, it undergoes de-shielding.}
\end{figure}

Chemical shift is a crucial parameter in NMR spectroscopy that unveils information about molecular structures and functional groups within the spectrum. Reflecting the proton environment, chemical shifts result from variations in energy absorption due to magnetic shielding. All protons are enveloped by electrons providing magnetic shielding from \(\boldsymbol{\vec{B_0}}\), and higher electron density enhances this shielding effect [see Fig. \ref{fig.2a}] \cite{sakurai1995modern, rabi1938new, bakhmutov2004nuclei}. The shielding is directly linked to the electron density around the \(^1\text{H}\) nucleus, with more shielded electrons requiring increased energy for resonance.

The effective magnetic field, denoted as $\boldsymbol{\vec{B}_{\text{eff}}}$, encountered by a nucleus, results from the combination of $\boldsymbol{\vec{B_0}}$ and the induced magnetic field, $\boldsymbol{\vec{B}_{\text{ind}}}$, which is generally of less magnitude than $\boldsymbol{\vec{B_0}}$. Mathematically, this relationship is expressed as
\begin{equation}
    \vec{B}_{\text{eff}} = \vec{B}_0 + \vec{B}_{\text{ind}} = \vec{B}_0(1 - \sigma) \label{eq:effective_magnetic_field}
\end{equation}

Where $\sigma$ denotes the shielding constant. In accordance with Lenz's Law, $\boldsymbol{\vec{B_0}}$ influencing the electron density of the $^1\text{H}$ proton opposes $\boldsymbol{\vec{B}_{\text{ind}}} (\vec{B}_{\text{ind}} = -\sigma\vec{B}_0)$. Nuclei in diverse local environments encounter slightly varying frequencies from $\omega_0$, termed the effective Larmor frequency $\omega_{\text{eff}}$ \cite{lancaster2014quantum, bakhmutov2004nuclei, shifrin1998new}.

The shielding constant for a specific nucleus in a given environment is expressed as
\begin{equation}
    \omega_{\text{eff}} = \gamma\vec{B}_{\text{eff}} = \gamma\vec{B}_0(1 - \sigma). \label{eq:shielding_constant}
\end{equation}

As illustrated in Fig. \ref{fig.2b}, when $\boldsymbol{\vec{B}_{\text{eff}}}$ at the nucleus position is less than $\boldsymbol{\vec{B_0}}$, the nucleus is shielded; otherwise, it is considered de-shielded. Due to electron shielding, the nucleus experiences distinct frequencies, known as the chemical shift ($\delta$), expressed in parts-per-million (ppm).

The chemical shift ($\delta_s$) of a sample with the resonance Larmor frequency ($\omega_s$) is independent of $\boldsymbol{\vec{B_0}}$ and depends on the nucleus's environment. By considering the resonance Larmor frequency of an internal reference ($\omega_{\text{ref}}$), the chemical shift of the sample in ppm can be calculated as:
\begin{equation}
    \delta_s = 10^6 \times \frac{\omega_s - \omega_{\text{ref}}}{\omega_{\text{ref}}}\label{eq:chemical_shift}
\end{equation}

\section{Bloch equations}

The measured signal in an NMR sample arises from the presence of net magnetization $\boldsymbol{\vec{M}}$, a concept initially proposed by Felix Bloch in 1946 \cite{bloch1946nuclear, purcell1946resonance, posener1959shape}. Fundamental to NMR technology is the relaxation of net magnetization, driven by spin-lattice and spin-spin relaxations within the system, causing the spins to decohere and return to their ground states \cite{maleki2021perfect, maleki2018recovery, maleki2017entanglement}.

\begin{figure}[h]
     \centering
     \begin{subfigure}[b]{0.35\linewidth}
         \centering   \includegraphics[width=\linewidth]{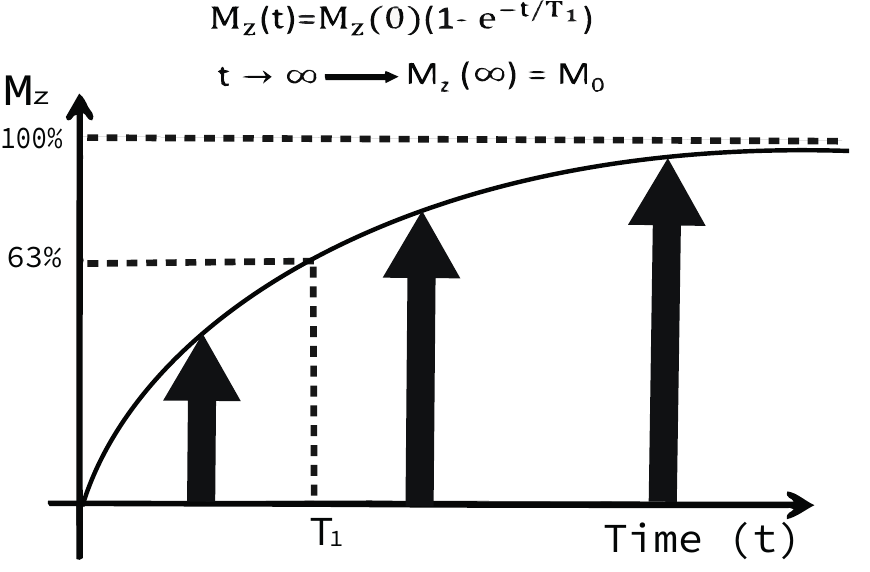}
         \caption{}
         \label{fig.3a}
     \end{subfigure}
     \quad
      \begin{subfigure}[b]{0.35\linewidth}
         \centering         \includegraphics[width=\linewidth]{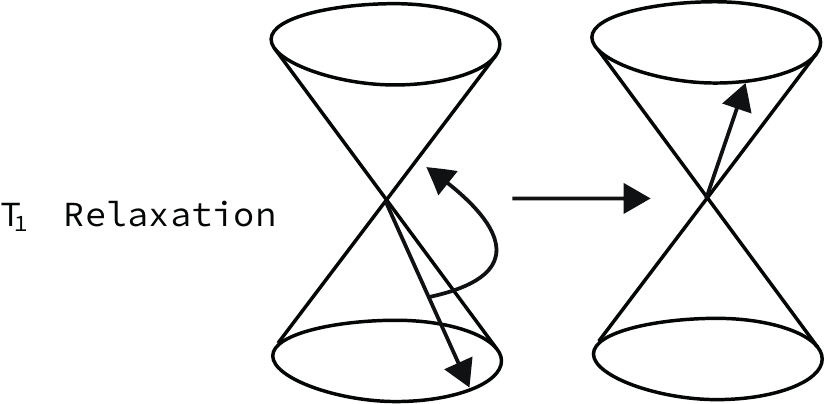}
         \caption{}
         \label{fig.3b}
     \end{subfigure}
     \quad
     \begin{subfigure}[b]{0.4\linewidth}
         \centering   \includegraphics[width=\linewidth]{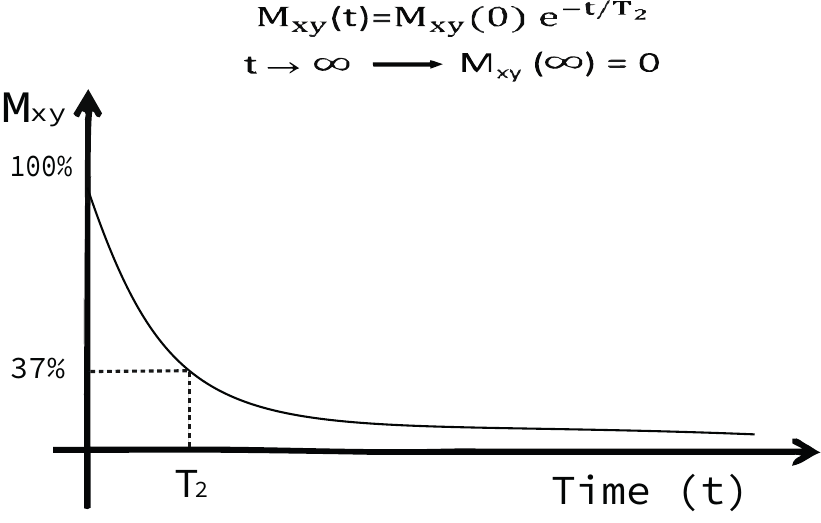}
         \caption{}
         \label{fig.3c}
     \end{subfigure}
      \quad
       \begin{subfigure}[b]{0.4\linewidth}
         \centering     \includegraphics[width=\linewidth]{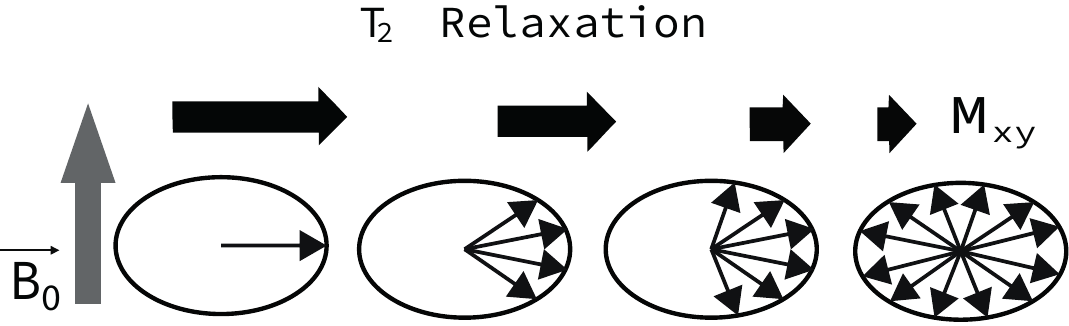}
         \caption{}
         \label{fig.3d}
     \end{subfigure}
      \quad \captionsetup{justification=Justified}
    \caption{\label{fig.3} (a)-(b) Application of an RF pulse results in the flipping of some spins to a higher energy state, causing a decrease in $M_z$. Upon turning off the RF pulse, $M_z$ gradually increases, ultimately reaching the initial magnetization $M_0$ at infinity. The point at which magnetization achieves approximately 63\% of $M_z$ is recorded as $T_1$. (c)-(d) Post RF pulse deactivation, the magnetization vector in the $x$-$y$ plane ($M_{xy}$) returns to its initial equilibrium, displaying maximum value conditions, and initiates precession at various Larmor frequencies. The time at which $M_{xy}$ reaches approximately 37\% of its maximum value is noted as $T_2$ relaxation time.}
\end{figure}

\textit{Spin-lattice relaxation}, or longitudinal relaxation, is the process where net magnetization returns to the $z$-axis, forming longitudinal magnetization ${M_z}$. During this phase, the net magnetization aligns with the external magnetic field \cite{stoneham1969shapes, vathyam1996homogeneous}. This relaxation, influenced by energy exchange between the spin system and neighboring molecules, leads to approximately 63\% of the magnetization returning to thermal equilibrium upon turning off RF pulses \cite{levy1975experimental, sepponen1985method}. Governed by the constant value ${T_1}$, an essential quantity in NMR spectroscopy, ${T_1}$ determines the rate at which the nucleus's spin becomes parallel to the magnetic field [see Fig. \ref{fig.3a}, \ref{fig.3b}] \cite{bakhmutov2004nuclei, shrivastava1983theory}.

\textit{Spin-spin relaxation}, or transverse relaxation, results from the interaction between the spin system and their magnetic field on the $x-y$ plane, forming transverse magnetization ${M_{xy}}$. Upon switching off an RF pulse, ${M_{xy}}$ decays to zero with the time constant ${T_2}$, leaving only around 37\% of the original $M_{xy}$ [see Fig. \ref{fig.3c}, \ref{fig.3d}] \cite{solomon1955relaxation, bushong2013magnetic, vathyam1996homogeneous, orbach1961spin}.

In the relaxation process, magnetization re-establishes equilibrium along the $z$-axis and in the $x-y$ plane at distinct rates. The Bloch equation describes the dynamic behavior of net magnetization as follows:
\begin{equation}
    \frac{d\vec{M}}{dt} = \vec{M} \times \gamma \vec{B} - \left[ \frac{M_x\hat{i} + M_y\hat{j}}{T_2} \right] - \left[ \frac{(M_z - M_0)\hat{k}}{T_1}\right]
\end{equation}
where $M_x$, $M_y$, and $M_z$ denote the magnetization components in the $x$, $y$, and $z$ directions, respectively, and $M_0$ represents the steady-state magnetization. Additionally, $T_1$ and $T_2$ stand for the spin-lattice and spin-spin relaxation times.

Examining each Bloch component in the magnetic field ($B_x$, $B_y$, $B_z$), we derive the following set of equations:
\begin{equation}
    \frac{dM_x(t)}{dt} = \gamma \left[M_y(t)B_z(t) - M_z(t)B_y(t)\right] - \frac{M_x(t)}{T_2}
\end{equation}
\begin{equation}
    \frac{dM_y(t)}{dt} = \gamma \left[ M_z(t)B_x(t) - M_x(t)B_z(t)\right] - \frac{M_y(t)}{T_2}
\end{equation}
\begin{equation}
    \frac{dM_z(t)}{dt} = \gamma \left[M_x(t)B_y(t) - M_y(t)B_x(t))\right] - \left[\frac{M_z(t) - M_0}{T_1}\right]
\end{equation}

\noindent At $B_x = 0$, $B_y = 0$, and $B_z = B_0$, the Bloch equation provides the following set of differential equations governing the rate of magnetization change:
\begin{equation}
    \frac{dM_x(t)}{dt} = \omega_0 M_y(t) - \frac{M_x(t)}{T_2} \label{eq:dMxdt}
\end{equation}
\begin{equation}
    \frac{dM_y(t)}{dt} = -\omega_0 M_x(t) - \frac{M_y(t)}{T_2} \label{eq:dMydt}
\end{equation}
\begin{equation}
    \frac{dM_z(t)}{dt} = \frac{M_0 - M_z(t)}{T_1} \label{eq:dMzdt}
\end{equation}
Equation \eqref{eq:dMzdt} corresponds to the Bloch equation for z-magnetization, governing the longitudinal component of magnetization \cite{jackson1999classical, shrivastava1983theory, tipler2003modern}. The rate of change of $z$-magnetization with time is inversely proportional to $T_1$ and directly proportional to the difference between the equilibrium magnetization value, $M_0$, and the $z$-magnetization at time $t$, $M_z(t)$ [see Fig. \ref{fig.3a}]. The solution for $M_z(t)$ can be simplified as:
\begin{equation}
    M_z(t) = M_z(0) ^{\left(-\frac{t}{T_1}\right)} + M_0 \left[1 - e^{\left(-\frac{t}{T_1}\right)}\right] \label{eq:Mz_solution}
\end{equation}
For $t = 0$, assuming $M_z(0) = 0$, Equation \eqref{eq:Mz_solution} further simplifies to:
\begin{equation}
    M_z(t) = M_0 \left[1 - e^{\left(-\frac{t}{T_1}\right)}\right] \label{eq:Mz_t0}
\end{equation}

\noindent Expressing that \(M_z(t)\) increases with time, converging exponentially toward the equilibrium value \(M_0\). Equation \eqref{eq:Mz_t0} illustrates that magnetization intensifies to approximately \(\left(1 - e^{-1}\right) \approx 63\%\) of its maximum within the time constant \(T_1\). In the limit where \(t \ll T_1\), the equation further simplifies to \(M_z(t) = (t/T_1)M_0\). Additionally, the longitudinal magnetization at \(t \rightarrow \infty\) eventually reaches its equilibrium value \(M_z(\infty) = M_0\).

\begin{figure}[h]
     \centering
     \begin{subfigure}[b]{0.3\linewidth}
         \centering   \includegraphics[width=\linewidth]{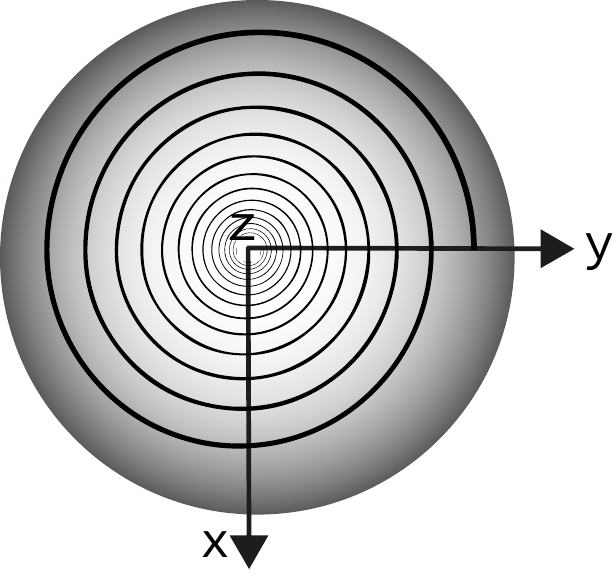}
         \caption{}
         \label{fig.4a}
     \end{subfigure}
      \begin{subfigure}[b]{0.4\linewidth}
         \centering
\includegraphics[width=\linewidth]{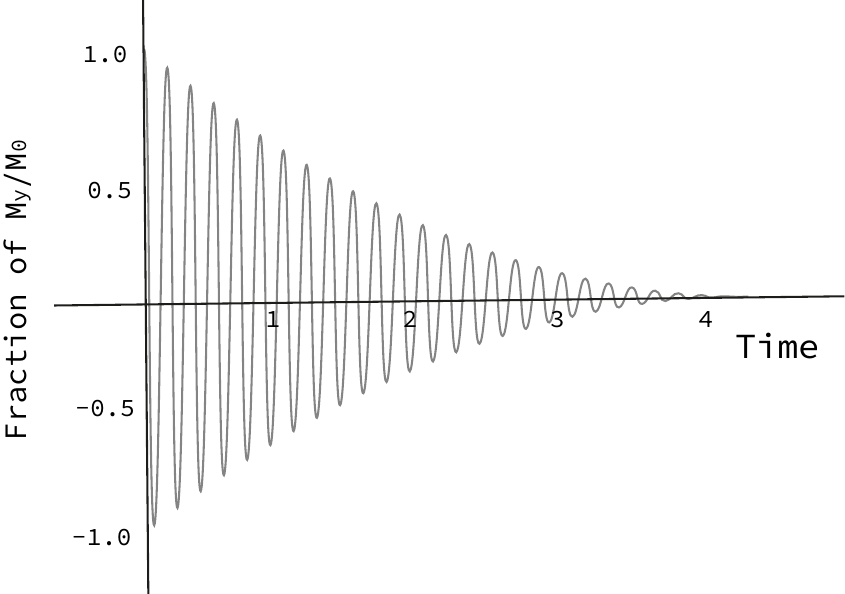}
         \caption{}
         \label{fig.4b}
     \end{subfigure}
     \begin{subfigure}[b]{0.3\linewidth}
         \centering
\includegraphics[width=\linewidth]{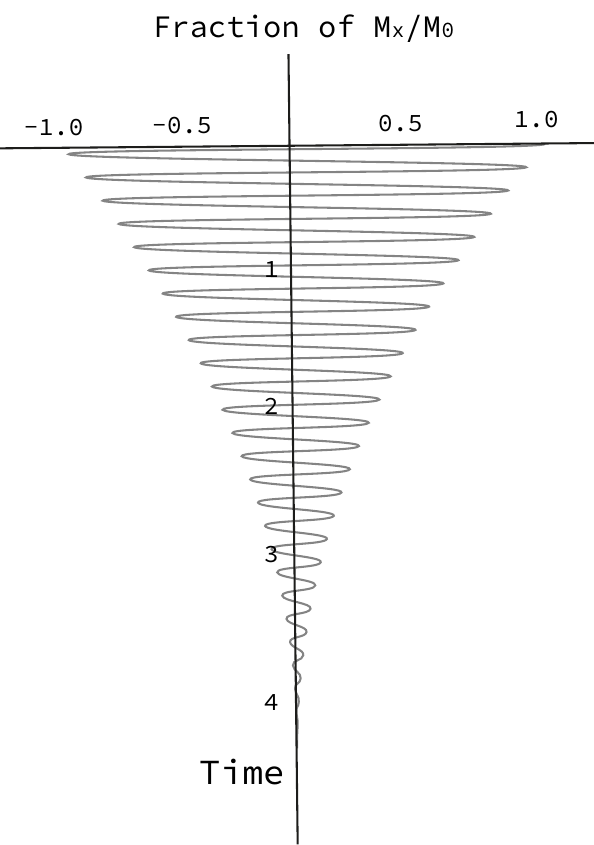}
         \caption{}
         \label{fig.4c}
     \end{subfigure}
       \begin{subfigure}[b]{0.35\linewidth}
         \centering    \includegraphics[width=\linewidth]{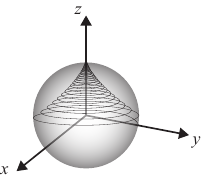}
         \caption{}
         \label{fig.4d}
     \end{subfigure}
      \captionsetup{justification=Justified}
    \caption{\label{fig.4} Upon application of an RF pulse, FID signal is acquired. Utilizing the FT method converts the FID signal into the NMR spectrum: (a) Depiction of transverse magnetization in the $x$-$y$ plane, where the external magnetic field aligns with the $z$-axis. (b)-(c) Representation of the FID signal along the $y$-axis and $x$-axis, respectively, their combination yielding (a). (d) The resultant NMR signal is presented in three dimensions.}
\end{figure}

The magnetization in the transverse plane is represented as a complex value, \(M_{xy}(t) = M_y(t) + iM_x(t)\). The solutions for \(M_{xy}(t)\) can be obtained by solving the differential equations provided by the Bloch equation:
\begin{align}
    M_{xy}(t) & = M_{xy}(0) e^{\left(-i\omega_0t\right)} e^{\left(-\frac{t}{T_2}\right)}\nonumber\\
    & = M_0 e^{\left(-i\omega_0t\right)} e^{\left(-\frac{t}{T_2}\right)} \nonumber\\
    & = M_0 \left(\cos(\omega_0t) - i\sin(\omega_0t)\right) e^{\left(-\frac{t}{T_2}\right)}
\label{eq:Mxy_solution}
\end{align}
It's noteworthy that $M_{xy}$ at $t = 0$ is defined as $M_{xy}(0) = M_0$. The real and imaginary parts of $M_{xy}(t)$, denoted by $\Re(M_{xy}(t))$ and $\Im(M_{xy}(t))$, respectively \cite{mcintyre2002spin, wang1986pictorial}, are given by:
\begin{equation}
    M_x(t) = \Im(M_{xy}(t)) = -M_0 \sin(\omega_0t) e^{\left(-\frac{t}{T_2}\right)}
\label{eq:Mx_solution}
\end{equation}
\begin{equation}
    M_y(t) = \Re(M_{xy}(t)) = M_0 \cos(\omega_0t) e^{\left(-\frac{t}{T_2}\right)} \label{eq:My_solution}
\end{equation}
Once $M_{xy}$ exists, magnetization quickly diminishes, as given by:
\begin{equation}
    M_{xy}(t) = M_0 e^{\left(-\frac{t}{T_2}\right)} \label{eq:Mxy_vanish}
\end{equation}
In the limit as $t \rightarrow \infty$, $M_{xy}$ vanishes, and $M_x(\infty) = M_y(\infty) = 0$ [see Fig. \ref{fig.3b}]. Therefore, the characteristic time \(T_2\) denotes the time required for the magnetization to decrease to \({1}/{e}\) of its initial value in the presence of the homogeneity field. A schematic illustrating the dynamics governed by the elements of the Bloch equation is presented in Figure \ref{fig.4}. This diagram offers a geometric representation of the relaxation process of NMR magnetization. The system's dynamics in the $x-y$ plane are depicted in Figure \ref{fig.4a}, showing the magnetization starting from the y-axis direction, progressing along the $z$-axis, and decaying to the center of the disc in the $x-y$ plane. This process is further explained from the perspectives of the $y$-axis and $x$-axis in Figures \ref{fig.4b} and \ref{fig.4c}, respectively. The complete dynamics, including the z-axis, can be visualized in Figure \ref{fig.4d}. The comparative dynamics of the $y$-axis and $x$-axis for the initial magnetization in the $y$-axis direction are illustrated in Figure \ref{fig.4c}.

Equation \eqref{eq:Mxy_vanish} is recognized as the Free Induction Decay (FID) signal, comprising decaying oscillation sine and cosine functions [refer to Figures \ref{fig.4a}, \ref{fig.4d}]. Through the Fourier Transformation (FT) method, denoted as \(F(\omega)\), the FID signal is transformed into the frequency domain, appearing as several peaks on the NMR spectrum. The FT method can be expressed as \(F(\omega) = \int_{-\infty}^{\infty} f(t) e^{(i\omega t)}dt\). \(F(\omega)\) is a complex function that can be separated into real ($\Re$)and imaginary ($\Im$) parts:

\begin{align}
    \Re(F(\omega)) = \int_{-\infty}^{+\infty} f(t) \cos(\omega t)dt, \nonumber &\\  \Im(F(\omega)) = \int_{-\infty}^{+\infty} f(t) \sin(\omega t)dt 
\end{align}

\section{Various NMR Spectroscopy Techniques}

NMR spectroscopy stands as a versatile technique for exploring a variety of atomic isotopes possessing a nuclear spin of \(1/2\), including \(^1H\), \(^{13}C\), \(^{31}P\), \(^{15}N\), and \(^{19}F\). In Table I, an exhaustive compilation is presented, enumerating diverse nuclei with a spin quantum number of \(1/2\), along with their corresponding gyromagnetic ratios (\(\gamma\)) and natural abundance percentages (\% Nat. Abd.) \cite{roberts1977basic, golman2006metabolic, filler2009history}. This section will provide an in-depth examination of NMR spectroscopy, focusing on the atomic isotopes delineated in Table \ref{tab:nuclei_properties}.

\begin{table}[htbp]
    \centering
    \caption{Properties of nuclei with a spin quantum number of \(1/2\). The table includes the natural abundance (\% Nat. Abd.) and gyromagnetic ratio (\(\gamma\)) for various isotopes.}
    \begin{tabular}{ccc}
        \hline
        \textrm{Nucleus} & \textrm{Nat. Abd (\%)} & \({\gamma}\) (\(T^{-1}s^{-1}\)) \\
        \hline
        \(^1\text{H}\) & 100.0 & \(2.6752 \times 10^8\) \\
        \(^{13}\text{C}\) & 1.11 & \(6.7283 \times 10^7\) \\
        \(^{19}\text{F}\) & 100.0 & \(2.517 \times 10^8\) \\
        \(^{15}\text{N}\) & 0.37 & \(-2.713 \times 10^7\) \\
        \(^{31}\text{P}\) & 100.0 & \(1.0841 \times 10^8\) \\
        \hline
    \end{tabular}
    \label{tab:nuclei_properties}
\end{table}

\subsection{Proton NMR Spectroscopy}
\textit{Proton} NMR or \textit{\(^1H\)}-NMR \textit{Spectroscopy:} Identifying distinct hydrogen atoms within a sample is facilitated by \(^1H\)-NMR, offering valuable insights into metabolic pathways within living cells. In this method, the biological sample is dissolved in a solvent serving as the NMR reference, devoid of protons, such as deuterium oxide (D\(_2\)O) [see Fig. \ref{fig.5a}] \cite{legchenko2002review, jung2011h}. The \(^1H\)-NMR spectrum exhibits peaks ranging from 0 to 14 ppm. Analysis of the NMR spectrum provides details on the number, positions, relative intensity, and splitting of signals \cite{levitt2013spin}. This information reveals the quantity of protons in the sample and the hydrogens responsible for generating the observed peaks. Additionally, it reflects the extent of signal interaction with neighboring hydrogen atoms, termed \textit{spin splitting} \cite{rugar2004single, westbrook2018mri}. The $n + 1$ rule governs the splitting of proton signals, with \(n\) representing the number of protons in the nearby nuclei. For instance, in the absence of nearby hydrogen atoms, the signal splitting, as per the $n + 1$ rule, results in a singlet. Various splitting types are illustrated in Fig. \ref{fig.6}. As indicated in Table \ref{tab:nuclei_properties}, \(^1H\)-NMR exhibits higher sensitivity compared to \(^{13}C\)-NMR due to its maximum natural abundance (100\%) and larger gyromagnetic ratio \(\gamma = 2.6752 \times 10^8 \, \text{T}^{-1}\text{s}^{-1}\) \cite{levitt1982broadband, slichter2013principles, krynicki1966proton}.
\subsection{Carbon NMR Spectroscopy}
\begin{figure}[h]
     \centering
     \begin{subfigure}[b]{0.27\linewidth}
         \centering         \includegraphics[width=\linewidth]{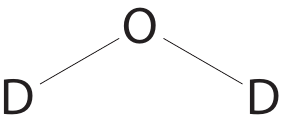}
         \caption{Deuterium oxide (D\textsubscript{2}O)}
         \label{fig.5a}
     \end{subfigure}
     \hspace{1cm}
      \begin{subfigure}[b]{0.3\linewidth}
         \centering   \includegraphics[width=\linewidth]{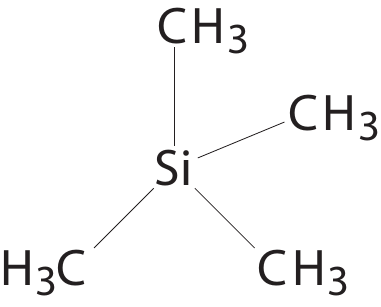}
         \caption{TMS (Si(CH\textsubscript{3})\textsubscript{4})}
         \label{fig.5b}
     \end{subfigure}
      \hspace{0.5cm}
     \begin{subfigure}[b]{0.3\linewidth}
         \centering
         \includegraphics[width=\linewidth]{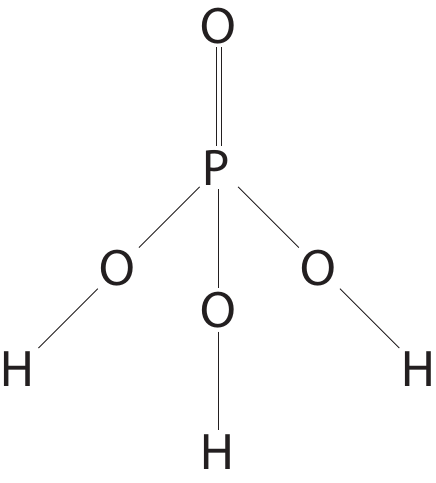}
         \caption{Phosphoric acid (H\textsubscript{3}PO\textsubscript{4})}
         \label{fig.5c}
     \end{subfigure}
       \hspace{0.5cm}
       \begin{subfigure}[b]{0.3\linewidth}
         \centering       \includegraphics[width=\linewidth]{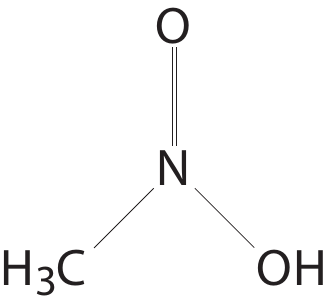}
          \hspace{0.5cm}
\caption{Nitromethane (CH\textsubscript{3}NO\textsubscript{2})}
         \label{fig.5d}
     \end{subfigure}
     \hspace{0.5cm}
       \begin{subfigure}[b]{0.23\linewidth}
         \centering
\includegraphics[width=\linewidth]{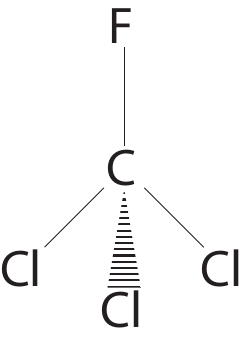}
         \caption{Trichlorofluoromethane (CFCl\textsubscript{3})}
         \label{fig.5e}
     \end{subfigure}
      \quad
      \captionsetup{justification=Justified}
    \caption{\label{fig.5} {Schematic representation of internal standard references used for various NMR types: (a) Deuterium oxide (D\textsubscript{2}O) for \textsuperscript{1}H-NMR, (b) TMS (Si(CH\textsubscript{3})\textsubscript{4}) for \textsuperscript{13}C-NMR, (c) Phosphoric acid (H\textsubscript{3}PO\textsubscript{4}) for \textsuperscript{31}P-NMR, (d) Nitromethane (CH\textsubscript{3}NO\textsubscript{2}) for \textsuperscript{15}N-NMR, and (e) Trichlorofluoromethane (CFCl\textsubscript{3}) for \textsuperscript{19}F-NMR.}}
\end{figure}

\begin{figure}[h]
    \centering    \includegraphics[width=0.6\linewidth]{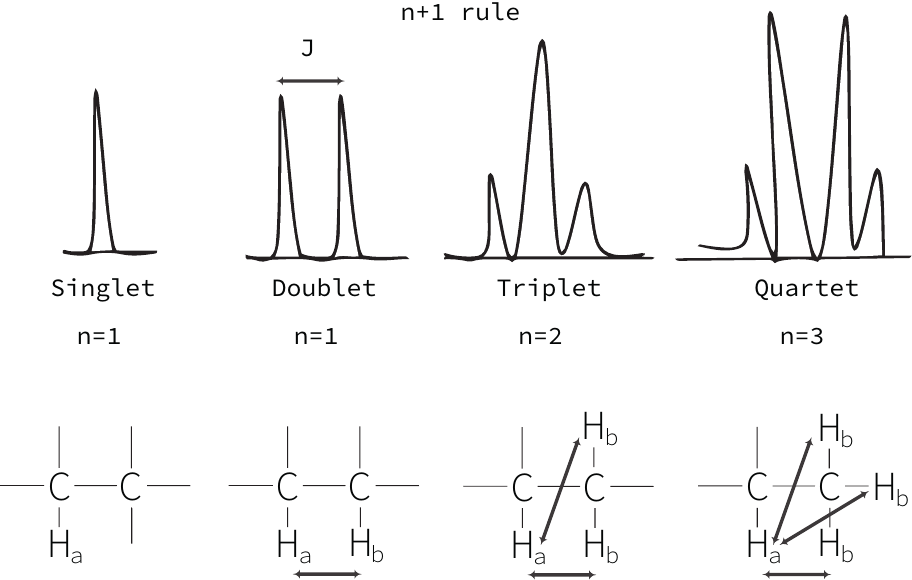}
\captionsetup{justification=Justified}
    \caption{Splitting patterns in NMR spectroscopy: The diversity of the signal is calculated by the $n+1$ rule. Singlet signals are not coupled to any protons; doublet signals are coupled to one proton; triplet signals are coupled to two protons; quartet signals are coupled to three protons. The coupling constant, J, represents the distance between the peaks in a splitting pattern. Illustrated with hydrogen nuclei H\textsubscript{a} and H\textsubscript{b}.}
    \label{fig.6}
\end{figure}

\textit{Carbon} NMR or \(^{13}C\)-NMR \textit{Spectroscopy:} proves to be highly valuable for the analysis of carbon-based chemical samples, encompassing all living systems and organic compounds. Only approximately 1.11\% of naturally occurring carbons possess a quantum spin number of \(1/2\) with \(\gamma = 6.7283 \times 10^7 \, \text{T}^{-1}\text{s}^{-1}\), endowing the \(^{13}C\) nucleus with activity in NMR imaging. However, \(^{13}C\)-NMR is less sensitive than \(^1H\)-NMR, as detailed in Table \ref{tab:nuclei_properties}. To obtain a reliable \(^{13}C\)-NMR spectrum, the sample should be mixed with a solvent serving as a reference compound, containing carbon, such as Tetramethylsilane (TMS: Si(CH\(_3\))\(_4\)), ensuring complete proton shielding [see Fig. \ref{fig.5b}] \cite{levitt2013spin, darbeau2006nuclear, grant1964carbon, schaefer1976carbon}.

In contrast to \(^1H\)-NMR spectra, \(^{13}C\)-NMR peaks span from 0 to 200 ppm, facilitating the detection of distinct peaks with a lower likelihood of signal overlapping in NMR spectroscopy \cite{graham1985anaerobic, soni2021brain, sonnewald1993metabolism}.

\subsection{Phosphorus NMR Spectroscopy}

The \(^{31}P\) nucleus boasts a 100\% natural abundance with \(\gamma = 1.0841 \times 10^8 \, \text{T}^{-1}\text{s}^{-1}\), rendering it akin to \(^1H\)-NMR. However, \(^{31}P\) chemical shifts exhibit a broader range in comparison to \(^1H\)-NMR. For the generation of a \(^{31}P\)-NMR signal, the NMR sample should be mixed with a solvent containing phosphorus, such as phosphoric acid (H\(_3\)PO\(_4\)) [refer to Fig. \ref{fig.5c}]. As indicated in Table \ref{tab:nuclei_properties}, \(^{31}P\)-NMR demonstrates lower sensitivity compared to \(^1H\)-NMR, with sharper peak than of \(^{13}C\)-NMR.

The combination of high natural abundance and a large \(\gamma\) makes \(^{31}P\)-NMR an ideal tool for investigating various systems, including the study of phospholipid liposomes \cite{dubinnyi2006modeling}, exploration of different food characteristics in food science, such as milk, green tea, and other foods \cite{gall2003nmr, renou1995nmr}. Moreover, it proves valuable in detecting cancer tumors in \textit{in-vivo} studies, such as breast cancer \cite{daly1987phospholipid} and lung cancer \cite{ng1982p}.

\subsection{Nitrogen NMR Spectroscopy}

\textit{Nitrogen} NMR or (\(^{15}N\)-NMR) \textit{Spectroscopy}: proves to be a potent tool for extracting valuable information from samples containing nitrogen atoms, such as RNA/DNA nucleobases \cite{barnwal2017applications}, the molecular structure of organic compounds \cite{von1986n}, and soil organic matter \cite{mathers2000recent}. As indicated in Table \ref{tab:nuclei_properties}, \(^{15}N\)-NMR exhibits significantly lower sensitivity compared to \(^1H\)-NMR and \(^{13}C\)-NMR, owing to the low natural abundance (37\%) of the \(^{15}N\) nucleus with its small \(\gamma\) value (\(\gamma = -2.713 \times 10^7 \, \text{T}^{-1}\text{s}^{-1}\)). Additionally, the \(^{15}N\)-NMR chemical shift is considerably broader than that of \(^{13}C\)- and \(^1H\)-NMR, spanning from 0 to 900 ppm. For NMR investigations with \(^{15}N\), nitromethane (\(\text{CH}_3\text{NO}_2\)) is utilized as a standard external reference, as depicted in Fig. \ref{fig.5e} \cite{khin1979some}.

\subsection{Fluorine NMR Spectroscopy}

\textit{Fluorine} NMR or (\(^{19}F\)-NMR) \textit{Spectroscopy}: proves to be an invaluable tool for analyzing various amino acids, nucleotides, and sugars containing fluorine. The \(^{19}\)F nucleus exhibits an approximate 100\% natural abundance with a substantial \( \gamma = 2.517 \times 10^8 \, \text{T}^{-1}\text{s}^{-1} \), providing high sensitivity and a broader chemical shift dispersion (ranging from -300 to 400 ppm) in comparison to 1H-NMR. Consequently, \(^{19}F\)-NMR is a convenient method for investigating essential compounds in fluorine-containing pharmaceuticals \cite{okaru2017application, yu2013new}. In \(^{19}F\)-NMR technique, trichlorofluoromethane (\(\text{C}\text{F}\text{Cl}_{3})\) is commonly utilized as a standard reference, and its structure is depicted in Fig. \ref{fig.5d}.

\section{Physics of MRI}

Medical Resonance Imaging or MRI, is a medical imaging technique based on NMR principles. This diagnostic method enables the observation of molecular fields and the acquisition of high-resolution images of soft tissues, such as the brain and cancer tumors \cite{plewes2012physics, bernstein2004handbook}. With the human body consisting of 70\% water containing hydrogen and oxygen atoms, and hydrogen nuclei being the most abundant, MRI allows the visualization of atomic nuclei within the body, producing robust signals \cite{filler2009history, hirsch2015brute}.

\begin{figure}[h]
    \centering
    \includegraphics[width=0.7\linewidth]{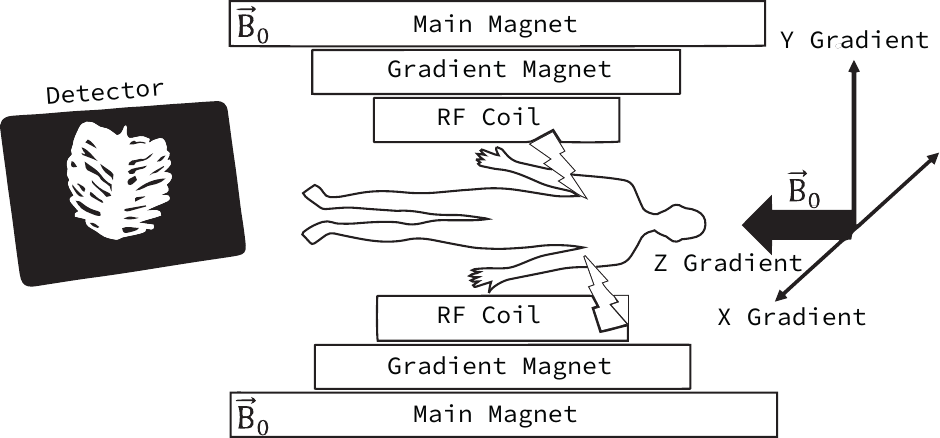}
    \captionsetup{justification=Justified}
    \caption{Schematic representation of MRI: The arrangement includes the primary magnet \(\vec{B_0}\) positioned along the $z$-axis, gradient coils distributed in the $x$-, $y$-, and $z$-axes, and RF receiver coils. The MRI system is linked to a computer for image processing of the body.}
    \label{fig.7}
\end{figure}

A key advantage of MRI lies in its high signal-to-noise ratio, attributed to the powerful magnetic field employed. MRI operates with a main magnetic field, typically ranging from 1.5 to 3T. According to the Zeeman effect, hydrogen atoms can align parallel or anti-parallel to \(\vec{B}_0\). The Boltzmann distribution dictates that more nuclei reside in the lower energy state than the higher energy state, leading to net magnetization [see Fig. \ref{fig.7}] \cite{canet1975time}. MRI machines also incorporate gradient magnets, constituting the second part of the MRI system. These magnets generate a magnetic field weaker than \(\vec{B}_0\) in the $x-$, $y-$, and $z-$axes, altering the strength of \(\vec{B}_0\) and enhancing precession frequencies through the slice-selection gradient. This process facilitates spatial encoding for MRI signals \cite{friston1994analysis, pruessmann1999sense}. Spatial encoding involves projecting signals from all slice spins along the gradient axes. By considering the spatial information obtained, axial images can be formed using data from the $Z-$ gradient along the long axis, coronal images can be created using data from the $Y-$ gradient along the vertical axis, and sagittal images can be generated from the $X-$ gradient along the horizontal axis \cite{friston1994analysis, okaru2017application, plewes2012physics}.

RF coils are also integrated into the MRI machine, transmitting RF pulses into the body and receiving signals to generate and display images. These coils are designed for specific body regions to improve the signal-to-noise ratio and enhance diagnostic image quality. In this context, an RF pulse \(\vec{B}_1\) orthogonal to \(\vec{B}_0\) and with the precession frequency \(\omega(x, y, z) = \gamma \vec{B}_0(x, y, z)\) is applied, turning the net magnetization toward the $x-y$ plane, as elaborated in detail for the NMR signal.

\begin{figure}[h]
    \centering
    \includegraphics[width=0.6\linewidth]{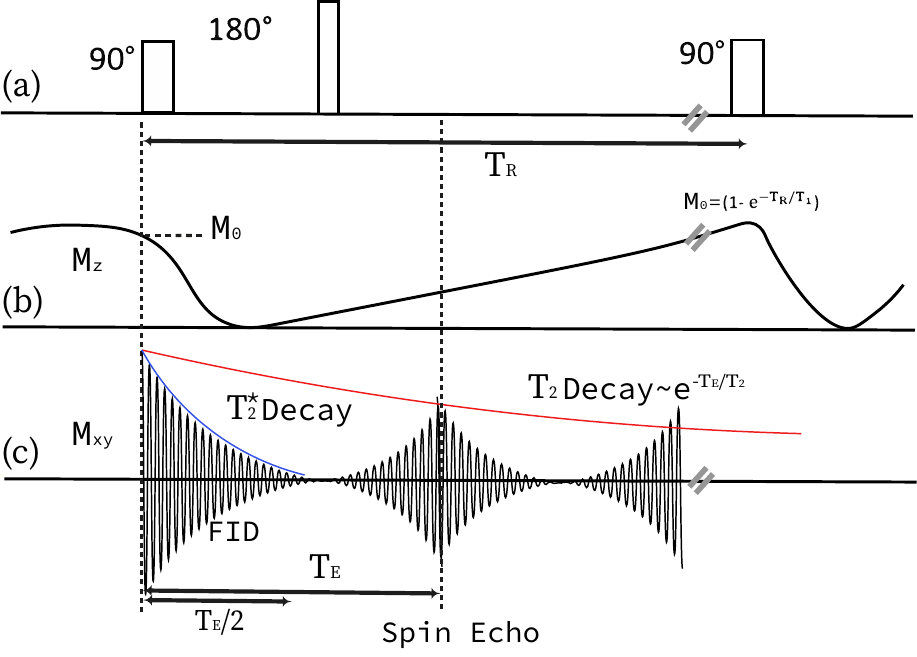}
    \captionsetup{justification=Justified}
    \caption{Pulse sequence scheme for generating the spin-echo signal: (a) Illustration of 90-degree RF pulses and a 180-degree pulse at the ($T_E/2$) point. (b) The longitudinal magnetization (\(M_z\)) decreases from the initial point, \(M_0\), and increases along the z-axis until the next 90-degree pulse (at the \(T_R\) point). (c) The transverse magnetization (\(M_{xy}\)) flips back into the $x-y$ plane, undergoes dephasing, and records \(T^*_2\) decays or the Free Induction Decay (FID) signal. The amplitude of the spin-echo signal is measured, and it depends on \(T_2\) at the \(T_E\) point. The spin echo process repeats after the \(T_R\) point.}
    \label{fig.8}
\end{figure}

\section{Spin Echo Pulse Sequence}

The spin-echo sequence plays a pivotal role in generating MRI images. By iteratively executing the pulse sequence, referred to as the spin-echo pulse sequence, high-quality images can be obtained \cite{yao2009application, mosher2004cartilage}. This sequence involves a 90-degree RF pulse followed by one or more 180-degree pulses to refocus the spins, as depicted in Fig. \ref{fig.8}.

Two crucial parameters characterize the spin-echo pulse sequence: the time between two 90-degree RF pulses, known as the repetition time ($T_R$), and the time between a 90-degree RF pulse and the formation of the echo signal in MRI ($T_E$), as shown in Fig. \ref{fig.8}a, \ref{fig.8}c. The magnitude of \(M_z\) at \(t = T_R\) is governed by:

\begin{equation}
    M_z(T_R) = M_0 \left[1 - e^{\left(-\frac{T_R}{T_1}\right)}\right] \label{eq:Mz_TR}
\end{equation}

The FID signal undergoes reversal and recovery by applying a 180-degree pulse at \(T_E/2\). Subsequently, magnetization relaxes along the \(-z\)-axis and returns to the \(+z\)-axis, ultimately reaching its original value \(M_0\). If a 90-degree RF pulse is applied during this process at any point , \(M_z\) rotates into the \(M_{xy}\) plane, facilitating signal observation \cite{basser1994estimation, antalek2002using}.

The dephasing of magnetization causes a decline in transverse net magnetization, which follows a decay characterized by \(T_2\). Subsequent dephasing after the echo leads to vector refocusing, giving rise to the formation of a second echo. This cyclic process can be iterated to generate multiple spin echoes. Following a 90-degree RF pulse, the expression for \(M_{xy}\) at \(T_E\) is given by:

\begin{equation}
    M_{xy}(T_E) = M_{xy}(0) e^{\left(-\frac{T_E}{T_2}\right)} \label{eq:Mxy_TE}
\end{equation}

\noindent The interplay of \(T_1\) and \(T_2\) in biological tissue determines the signal output, \(S(r)\), defined by:
\begin{equation}
    S(r) \approx \rho \left[1 - e^{\left(-\frac{T_R}{T_1}\right)}\right] e^{\left(-\frac{T_E}{T_2}\right)} \label{eq:signal_output}
\end{equation}
Here, \(\rho\) denotes the density of protons per unit tissue. A higher proton density enhances image contrast, resulting in a more luminous signal on the screen and improving the overall quality of the MRI signal \cite{styner2000parametric, chang1992technique, vovk2007review}.

Another concept related to transverse relaxation is \(T_2^*\) decay, occurring between a 90-degree pulse and \(T_E/2\) [see Fig. \ref{fig.8}c]. In the field's inhomogeneity, denoted by \(\Delta B(r)\) and assumed time-independent, the spin system precesses with a Larmor frequency \(\omega(r) = \gamma\Delta B(r)\), causing \(M_{xy}(r, t)\) to decay as:
\begin{equation}
    M_{xy}(r, t) = M_{xy}(r, 0) \exp\left(-i\gamma\Delta B(r)t\right)
    \label{eq:Mxy_decay}
\end{equation}

The \(T_2^*\) characterizes the decay in the gradient echo (GRE) technique, significantly reducing scan time. The relationship between \(T_2\) and \(T_2^*\) is expressed as
\begin{equation}
    \frac{1}{T_2^*} = \frac{1}{T_2} + \frac{1}{T_{2i}} \label{eq:T2_star_relation}
\end{equation}
Here, \(T_2\) accounts for signal decays due to spin-spin relaxation, while \(T_{2i}\) represents transverse relaxation due to static magnetic field inhomogeneities that lead to coherence loss. The \(T_2^*\) decays are consistently less than or equal to \(T_2\) decays.

\section{Properties of Low-Field NMR and MRI}

The pursuit of high-resolution MRI has spurred researchers to explore innovative techniques employing high magnetic fields. While high magnetic field MRI is crucial to address the challenge of low magnetic moments in nuclear spins, advancements in MRI techniques have also made low-field MRI feasible. A significant advantage of low-field MRI is its provision as a cost-effective alternative to high-field instruments for MR imaging.

Furthermore, the operational frequency is much lower for low-field systems. For example, a 0.06T system operates at approximately 2.46 MHz, whereas a 3T system operates closer to 128 MHz. Additionally, the magnetic field provided at the lower field is notably homogeneous, making it an excellent resource for clinical research \cite{griffiths2018introduction, albert1999t1, maxwell1864dynamical}.

One limitation of low-field MRI is its sensitivity to noise originating from the environment, known as external electromagnetic interference (EMI) signals. Various developments have been introduced to mitigate and eliminate EMI during the scanning process. Another aspect of low-field MRI involves extending relaxation time using a contrast agent \cite{feynman2011feynman, bardeen1957theory}.

It is worth noting that certain MRI advancements aim to enhance the Signal-to-Noise Ratio (SNR) value in low-field MRI. Hyperpolarized (HP) MRI, utilizing dynamic nuclear polarization (DNP), enhances the polarization of spins in an MRI sample compared to the thermal equilibrium state. HP MRI is more efficiently performed at low fields, as the spins relax back to their equilibrium state more slowly.

\section{Understanding the Implications of the Heisenberg Uncertainty Principle}

In NMR and MRI, the energy operator involved in spectroscopy is denoted as the energy transition, with the corresponding operator termed the Hamiltonian (\(H\)). Analyzing the system involves obtaining its eigenvectors and eigenvalues. The presentation of energy values for a nucleus requires examining the wave function as a sum of various states aligned with or against the field, featuring probability distributions contingent on system characteristics \cite{abragam1983principles, schrodinger1926undulatory}. The Heisenberg time-energy uncertainty principle sheds light on the spectral width (\(SW\)) of the NMR spectrum \cite{sakurai1995modern, khashami2024fundamentals, rabi1938new, plewes2012physics}. Expressing the uncertainty principle for the nucleus in terms of frequency (\(\Delta\nu\)) and energy dispersion (\(\Delta E\)), we have

\begin{equation}
    \Delta t \Delta E \approx \Delta t(h\Delta\nu) \approx h \label{eq:time_energy_uncertainty}
\end{equation}

Furthermore, the nucleus remains in a state for an uncertain lifetime (\(\Delta t\)). The peak width (in Hz) for the nucleus, relaxing back to the lower state with a time constant (\(T_2\)) at frequency (\(\nu\)), can be expressed as
\begin{equation}
    \Delta\nu \approx \frac{1}{\Delta t} \approx \frac{1}{T_2} \label{eq:peak_width}
\end{equation}
Sharper peaks can be detected by recording a longer lifetime (\(\Delta t\)) or increasing the relaxation time (\(T_2\)), resulting in a reduced resonance linewidth (\(\Delta\nu\)). In this context, \(\Delta\nu\) denotes the range of frequencies affected by the lifetime (\(\Delta t\)).

\section{Applications of NMR and MRI in Medical Science}

\begin{figure}[h]
     \centering
     \begin{subfigure}[b]{0.45\linewidth}
         \centering   \includegraphics[width=\linewidth]{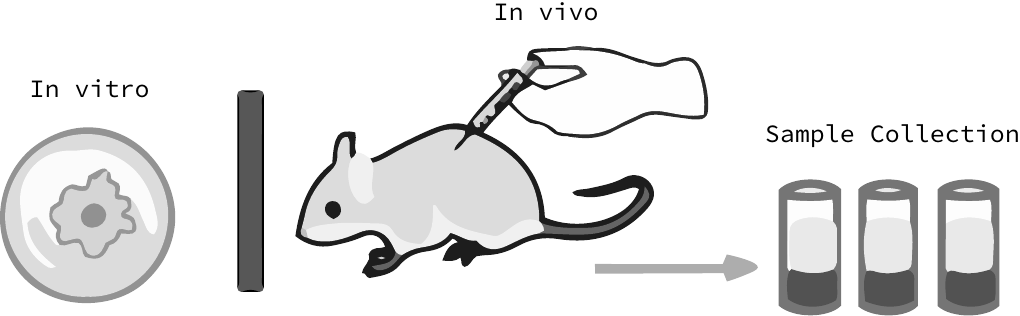}
         \caption{}
         \label{fig.9a}
     \end{subfigure}
     \quad
      \begin{subfigure}[b]{0.4\linewidth}
         \centering    \includegraphics[width=\linewidth]{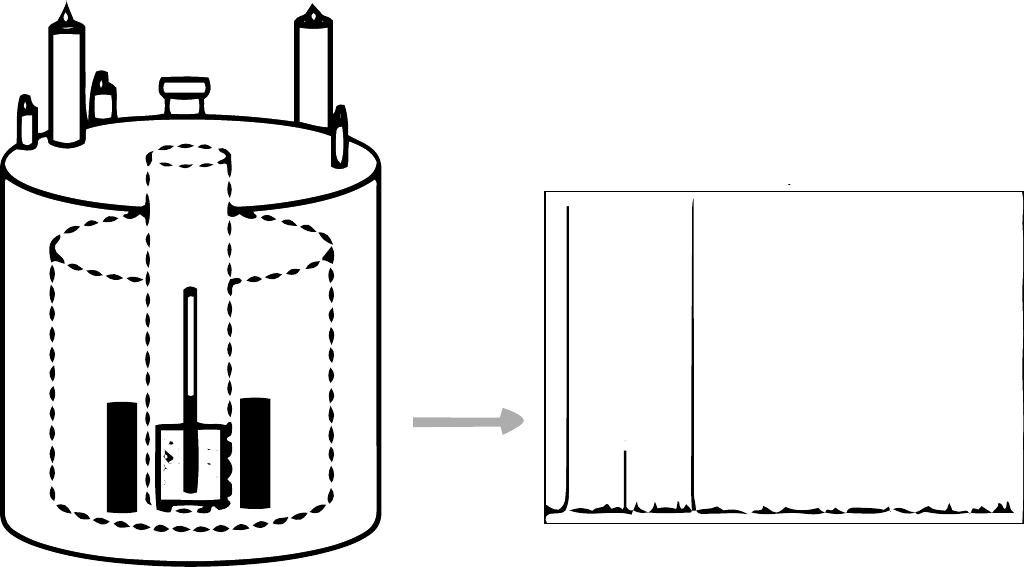}
         \caption{}
         \label{fig.9b}
     \end{subfigure}
     \quad
     \begin{subfigure}[b]{0.4\linewidth}
         \centering
    \includegraphics[width=\linewidth]{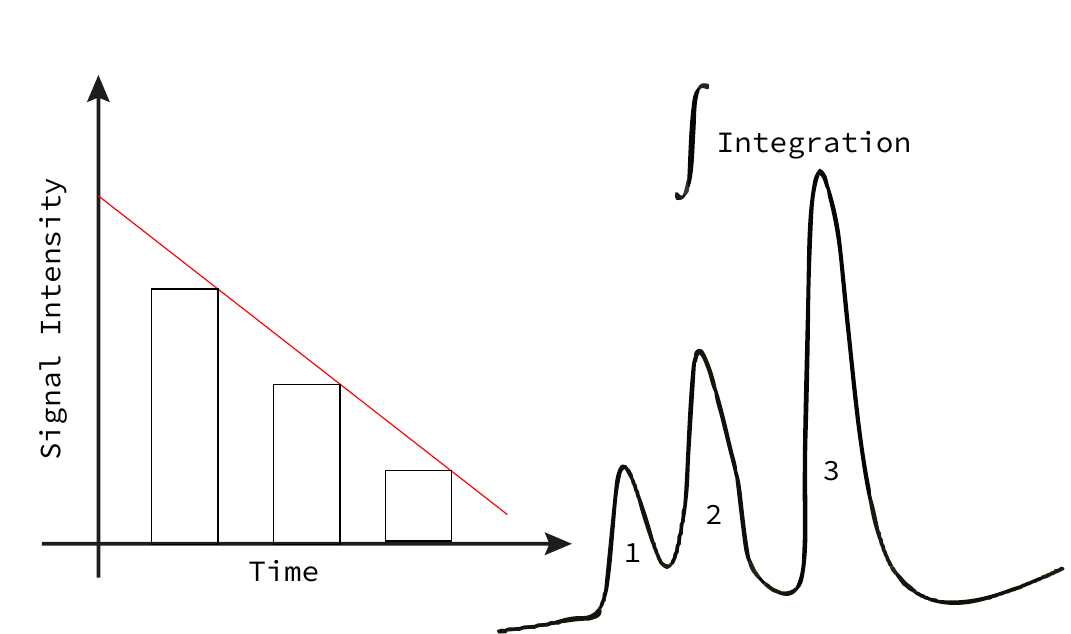}
         \caption{}
         \label{fig.9c}
     \end{subfigure}
      \quad
       \begin{subfigure}[b]{0.4\linewidth}
         \centering \includegraphics[width=\linewidth]{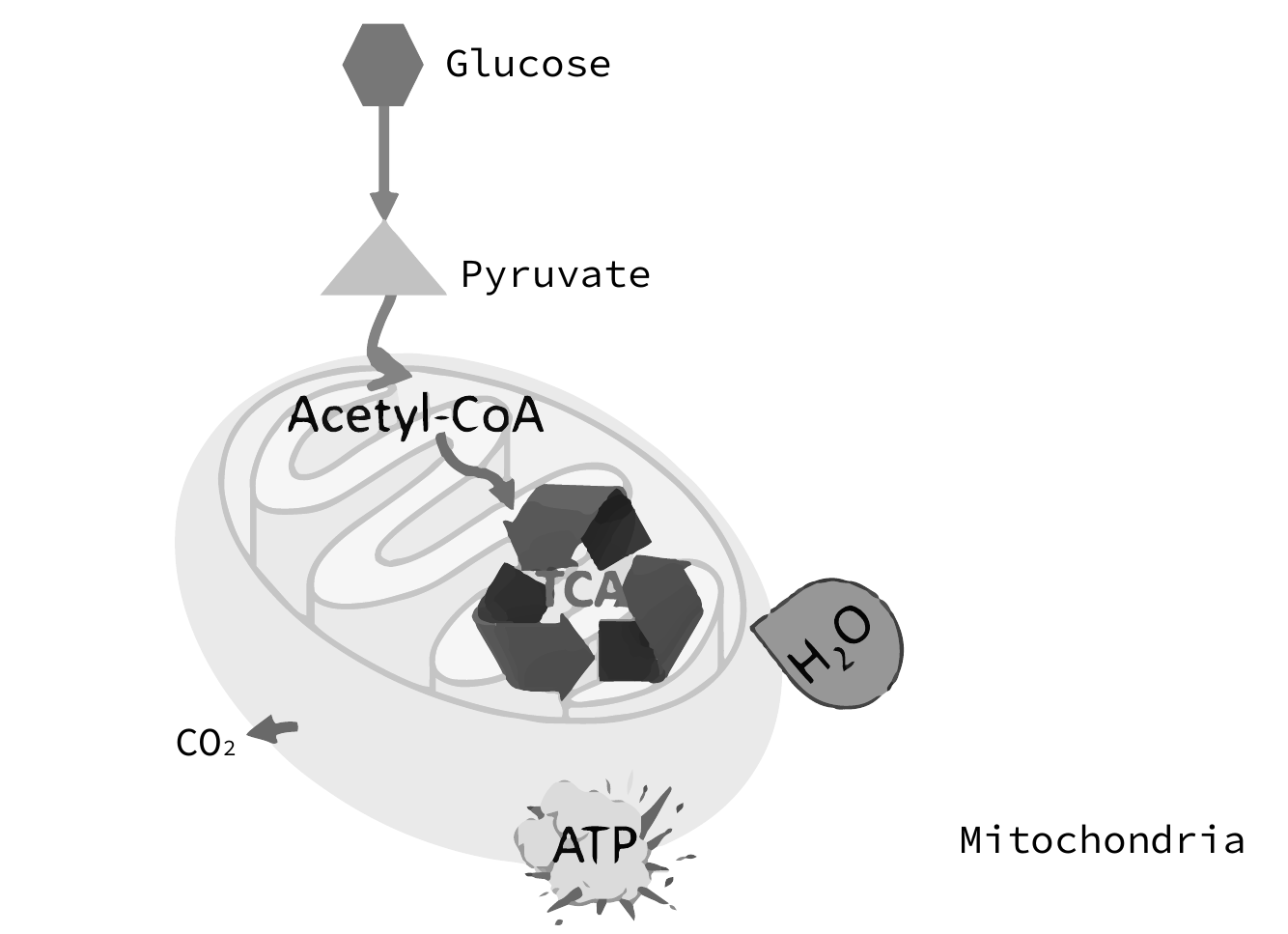}
         \caption{}
         \label{fig.9d}
     \end{subfigure}
      \quad
      \captionsetup{justification=Justified}
    \caption{\label{fig.9} Fundamentals of NMR data collection: (a) Preparation of the sample, (b) Detection of the NMR signal, (c) Analysis of the data, and (d) Representation of the metabolic pathway.}
\end{figure}

The NMR or MRI samples are obtained from experiments conducted either in cell culture (\textit{in-vitro}) or in animal and human models (\textit{in-vivo}), as illustrated in Fig. \ref{fig.9a}. In the in-vitro setup, isotope molecules like \(^1\text{H}\), \(^{13}\text{C}\), \(^{31}\text{P}\), \(^{15}\text{N}\), and \(^{19}\text{F}\) are dissolved in specific cell culture media and added to cell culture dishes to monitor their metabolism. As previously mentioned, each isotope molecule has specific standard external references used to normalize the output signal from the scanner. Once the powder is introduced into the cell culture dishes, the obtained cell sample is then transferred into the NMR tube and positioned within the magnet. Following this, the resulting NMR spectrum is processed and presented on a monitor for in-depth metabolic analysis, as depicted in Fig. \ref{fig.9b}. In the case of in-vivo studies, the sample is prepared and administered into the body's soft tissue for subsequent analysis.

The objective of the NMR technique is to accurately identify and quantify metabolic alterations in cells or organs. In the context of MR imaging, the sample is prepared accordingly, and an MRI scanner captures images of the subject. Subsequently, the MRI files are stored in the Digital Imaging and Communications in Medicine (DICOM) format, ready for subsequent analysis, as depicted in Fig. \ref{fig.9c}.

To conduct accurate data analysis, a comprehensive understanding of the metabolic pathways and general physiological roles of cells is essential. As illustrated in Fig. \ref{fig.9d}, glucose metabolism serves as the primary energy source for most organs. Initially, complex glucose molecules are transported into cell membranes via dedicated transporters. Within the cell, enzymes degrade glucose into more straightforward compounds, producing adenosine triphosphate (ATP), carbon dioxide (CO\(_2\)), and water (H\(_2\)O) within the mitochondria. The mitochondria play a vital role in this metabolic pathway. As the concluding product of the glycolysis pathway, pyruvate, originating from glycolysis, traverses the citric acid cycle or TCA, contributing to the oxidative phosphorylation process that yields ATP.

\section{Conclusion}

NMR and MRI stand out as versatile tools applicable across various fields, ranging from physics and chemistry to geology and medical science studies. This review has delved into the foundational principles of NMR and MRI technologies, tracing their roots from quantum mechanics to their crucial role in medical science. Commencing with a quantum mechanical foundation, we emphasized the substantial significance of NMR and MRI in advancing clinical research. Additionally, we provided a brief overview of different NMR systems and explored key applications of MRI techniques, offering valuable methods for visualizing internal structures within the body and soft tissues.




\section*{Declarations}

\subsubsection*{Ethical Statement}
As this manuscript is a comprehensive review and does not involve original research with human or animal subjects, ethical approval, and informed consent are not applicable. The study adheres to the principles and standards outlined for literature reviews. No funding was received for this review, and there are no conflicts of interest to declare.

\begin{itemize}
\item Funding: NA
\item Conflict of interest: NA
\item Ethics approval: NA
\item Consent to participate: NA
\item Data availability: NA
\item Authors' contributions: Islam G. Ali is the sole author of this review article.
\end{itemize}
\bibliography{main}

\end{document}